\documentclass[english,twoside,a4paper,10pt]{article}

\usepackage[latin1]{inputenc}
\usepackage[T1]{fontenc}
\usepackage{amsmath}
\usepackage{amsfonts}
\usepackage{graphicx}
\usepackage{subfigure}
\usepackage{a4wide}
\usepackage{amssymb}
\usepackage{fancyhdr}
\usepackage{mathrsfs}
\usepackage[toc,page]{appendix}

\begin{document}

\title{\bf $k$-essence in Horndeski models}
\author{ 
Shynaray Myrzakul\footnote{Email: shynaray1981@gmail.com},\,\,\,
Ratbay Myrzakulov\footnote{Email: rmyrzakulov@gmail.com},\,\,\,
Lorenzo Sebastiani\footnote{E-mail: lorenzo.sebastiani@unitn.it
}\\
\\
\begin{small}
Department of General \& Theoretical Physics and Eurasian Center for
\end{small}\\
\begin{small} 
Theoretical Physics, Eurasian National University, Astana 010008, Kazakhstan
\end{small}\\
}

\date{}

\maketitle


\begin{abstract}
In this paper, we investigate a simple class of Horndeski models where the scalar field plays the role of a $k$-essence fluid. We present several solutions for early-time universe, namely inflation and cosmological bounce, by making use of some reconstruction technique. Moreover, we furnish the formalism to calculate perturbations in FRW space-time and we compute the spectral index and the tensor-to-scalar ratio during inflation. 
\end{abstract}



\tableofcontents
\section{Introduction}

The interest in modified theories of gravity has considerably grown up in the last years, due to the possibility to reproduce with their wide choice of models a huge variety of cosmological scenarios
(see Refs.~\cite{R1, R2, R3, R4, R5} for some reviews). The field equations of modified gravity are much more involved with respect to the ones of General Relativity, but in 1974 Horndeski found a class of scalar tensor theories where the 
equations of motion appear at the second order like in the theory of Einstein~\cite{Horn}, making their investigation quite simple despite the complexity of the Lagrangian. A natural application of this kind of theories is related to the early-time acceleration that universe underwent after the Big Bang, namely the inflation~\cite{Linde, revinflazione}. In inflationary cosmology, a scalar field subjected to some suitable potential drives the accelerated expansion. In Horndeski models, the scalar field is coupled with gravity and higher curvature terms appear in the Lagrangian. Since it is expected that in hot universe scenario higher curvature corrections (motivated by quantum effects or string theories) emerge in the theory of gravitation, the studies of inflation in Horndeski gravity are popular and have been carried out in many works~\cite{Def, DeFelice, Kob, Kob2, Qiu, Maselli, BambaGal, EugeniaH, mioH, cognoH, add1, add2, add3, add4, addlast}. 

In this paper, we will consider a simple class of Horndeski models where the scalar field represents a $k$-essence fluid~\cite{kess1, kess2, kess3}, whose Lagrangian contains non-standard higher order kinetic term (see also Ref.~\cite{Bab}). 
We should note that
$k$-essence is motivated by string theory and offers a valid alternative description with respect to the standard canonical scalar field, since it brings to the suppression of the sound speed and therefore to an extremely small value of tensor-to-scalar ratio which is strongly favoured by observations.
Several solutions for early-time universe, namely inflation and cosmological bounce, will be investigated. We will see how the Horndeski contribution to the Lagrangian 
influences the perturbations left at the end of inflation. The perturbations reveal the short scale of initial conditions of the universe and must be in agreement with cosmological data.
We also observe that quantum gravity corrections may include all quadratic higher-derivative corrections~\cite{buch}, what leads to very interesting quantum gravity induced inflation (see Refs.~\cite{quattro, cinque, sei, sette}).

The paper is organized in the following way. In Section {\bf 2}, we will present the model and we will derive the equations of motion on flat Friedmann-Robertson-Walker metric.  Section {\bf 3} is devoted to the study of inflation. Some forms of $k$-essence models will be considered and a reconstructive method by starting from the solutions will be discussed. In Section {\bf 4} the cosmological bounce is investigated. In Section {\bf 5} the perturbations during inflation will be analyzed and the spectral index and the tensor-to-scalar ratio for our theory will be derived. Conclusions and final remarks are given in Section {\bf 6}.

We use units of $k_{\mathrm{B}} = c = \hbar = 1$ and 
$8\pi/M_{Pl}^2=1$, where $M_{Pl}$ is the Planck Mass.

\section{The model}

Horndeski Lagrangian~\cite{Horn} collects the most general class of scalar-tensor models with field equations at the second order like in General Relativity (GR),
\begin{equation}
I=\int_\mathcal M dx^4\sqrt{-g}\left[\frac{R}{2}+\mathcal L_H+L_m\right]\,,\quad \mathcal{L}_H=\sum_{i=2}^5\mathcal{L}_i\,,\label{action01}
\end{equation}
with
\begin{equation}
\mathcal{L}_2=P(\phi,X)\,,\nonumber
\end{equation}
\begin{equation}
\mathcal{L}_3=-G_3(\phi,X)\Box\phi\,,\nonumber
\end{equation}
\begin{equation}
\mathcal{L}_4=G_4(\phi,X)R+G_{4,X}[(\Box\phi)^2-(\nabla_\mu \nabla_\nu \phi)(\nabla^\mu \nabla^\nu \phi)]\,,\nonumber
\end{equation}
\begin{eqnarray}
\mathcal{L}_5&=&G_5(\phi,X)G_{\mu\nu}(\nabla^\mu \nabla^\nu \phi)-\frac{1}{6}G_{5,X}[(\Box\phi)^3-\nonumber\\&&
3(\Box\phi)(\nabla_\mu \nabla_\nu \phi)(\nabla^\mu \nabla^\nu \phi)+2(\nabla^\mu \nabla_\alpha\phi)(\nabla^\alpha \nabla_\beta \phi)(\nabla^\beta \nabla_\mu \phi)]\,.
\label{cGR}
\end{eqnarray}
In Eq.~(\ref{action01}), $\mathcal{M}$ represents the space-time manifold, $g$ is the determinant of the metric tensor $g_{\mu\nu}$, $R$ is the Ricci scalar of the Hilbert-Einstein action of GR, $\mathcal L_m$ is the Lagrangian of the matter contents of the space-time, while $L_H$ includes the higher curvature corrections to GR expressed by (\ref{cGR}), where we see that a scalar field is coupled with gravity. Here,
$G_{\mu\nu}:=R_{\mu\nu}-R g_{\mu\nu}/2$ is the usual Einstein's tensor, $R_{\mu\nu}$ being the Ricci tensor, while
$P(\phi, X)$ and $G_i(\phi, X)$ with $i=3,4,5$ are functions of the scalar field $\phi$ and its kinetic energy $X=-g^{\mu\nu}\partial_\mu\phi\partial_\nu\phi/2$.

We will consider the following simplified subclass of Horndeski models in absence of matter:
\begin{equation}
I=\int_\mathcal{M}d^4x\sqrt{-g}\left[\frac{R}{2}+P(\phi, X)\right]+I_H\,,\label{action}
\end{equation}
where\footnote{For a comparison with Ref.~\cite{DeFelice}, we have to set 
$G_3=\beta\phi$, $G_4=\alpha X$, $G_5=\gamma\phi$ with $X=\dot\phi^2/2$.}
\begin{equation}
I_H=\int_{\mathcal M} d^4 x \sqrt{-g}\left[\alpha\left(
G_{\mu\nu}\nabla^\mu\phi\nabla^\nu\phi\right)
+
\gamma 
\phi G_{\mu \nu} \nabla^\mu \nabla^\nu \phi -\beta \phi  \Box \phi \right]\,,\label{actionH}
\end{equation}
with $\alpha\,,\beta\,,\gamma$ constants.
We observe that after integration by part we also have,
\begin{equation}
\int_\mathcal{M}d^4x\sqrt{-g} \phi G_{\mu\nu} \nabla^\mu \nabla^\nu \phi=
\int_\mathcal{M}d^4x\sqrt{-g}\left( -\frac{g^{\mu\nu}\partial_\mu\phi\partial_\nu\phi R}{2} +(\Box \phi)^2-\nabla_\mu \nabla_\nu \phi \nabla^\mu \nabla^\nu \phi\right)\,,
\end{equation}
and in the spacial case $\alpha=\gamma$ we obtain a total derivative and the corresponding contributes disappear from the field equations. 
This theory is rich of cosmological applications (see Refs.~\cite{EugeniaH,mioH}) and in this paper we will identify $\phi$ with a $k$-essence field whose stress energy-tensor is given by~\cite{kess1, kess2},
\begin{equation}
T^{\mu}_{(\phi)\nu}=(\rho(\phi, X)+p(\phi, X))u^{\mu}u_{\nu}+p(\phi, X)\delta^{\mu}_\nu\,,\quad u_\nu=\frac{\partial_\nu\phi}{\sqrt{2X}}\,,\label{Tfluid}
\end{equation}
such that $P(\phi, X)\equiv p(\phi, X)$ corresponds to the effective pressure of $k$-essence and $\rho(\phi, X)$ to its energy density, due to the fact that the variation respect to the metric leads to
\begin{equation}
\rho(\phi, X)=2X\frac{\partial p(\phi, X)}{\partial X}-p(\phi, X)\,.\label{rhophi}
\end{equation}
For canonical scalar filed one has $p(\phi, X)=X-V(\phi)$, $V(\phi)$ being a function of the field only, but in general the Lagrangian of $k$-essence contains higher order kinetic term.

We will work in flat Friedmann-Robertson-Walker (FRW) space-time,
\begin{equation}
ds^2=-dt^2+a(t)^2 d{\bf x}^2\,,\label{metric}
\end{equation}
where $a\equiv a(t)$ is the scale factor depending on the cosmological time. We immediately have
\begin{equation}
X=\frac{\dot\phi^2}{2}\,,
\end{equation}
and the equations of motion (EOMs) read
\begin{eqnarray}
3H^2(1-3\alpha\dot\phi^2+3\gamma\dot\phi^2)=\rho(\phi, X)-\beta\dot\phi^2\,,\label{EOM1}
\end{eqnarray}
\begin{eqnarray}
-(2\dot H+3H^2)&=&p(\phi, X)-\beta\dot\phi
+\alpha\dot\phi^2(3H^2+2\dot H)
-6\alpha H^2\dot\phi^2-4\alpha H\dot\phi\ddot\phi-4\alpha\dot H\dot\phi^2\nonumber\\&&
+\gamma(2\dot H\dot\phi^2+4H\dot\phi\ddot\phi+3H^2\dot\phi^2)\,,\label{EOM2}
\end{eqnarray}
the dot being the derivative with respect to the time.
Finally, the continuity equation of $k$-essence coupled with gravity is given by
\begin{eqnarray}
&&
\dot\rho(\phi, X)+3H(\rho(\phi,X)+p(\phi, X))=\nonumber\\&&
-\ddot\phi\dot\phi(-2\beta+6\alpha H^2-6\gamma H^2)
-3H\dot\phi^2(-2\beta+6\alpha H^2-6\gamma H^2)
-12H\dot H\dot\phi^2(\alpha -\gamma)\,,\label{conslaw}
\end{eqnarray}
where $\rho(\phi, X)+p(\phi, X)=2X p_X(\phi, X)$.

\section{Models for inflation}

The considered simplified subclass of Horndeski models (\ref{action})--(\ref{actionH}) presents a sufficiently quite involved Lagrangian with the account of several higher curvature corrections to Einstein's gravity. It is expected that 
such a kind of terms
(maybe related to quantum effects) modify the theory of Einstein at high energy, when inflation takes place. In our case, the early-time de Sitter expansion is supported
by $k$-essence, which is a valid alternative description with respect to the standard canonical scalar field of ``old inflationary scenario''. We should note that $k$-essence is strictly connected with string theory and permits to suppress the sound speed leading to a negligible tensor-to-scalar ratio according with cosmological observations (see \S\ref{pert}). 
On the other side, the curvature corrections coupled with the $k$-field contribute to the graceful exit from inflation. In fact,
posing that $\dot\phi^2=H^2 x^2$, in the FRW-field equations
 we aquire additional terms proportional to $\sim H^4$. During inflation $x^2\ll 1$ (in Planck units)
and this terms 
appear in the perturbations at the origin of the anisotropies in our universe.
In this respect, despite the arena of models for inflation is in principle infinite, the accuracy of cosmological data are offering compelling bounds for deviations of Einstein's gravity and discriminate between viable and not viable models.  Thus, the introduction of additional
extra-degrees of freedom renders the theory more flexible and may help to reach at least some intermediate results in the study
of the primordial phases of the expansion of our universe.

The evolution of inflationary universe is described by the $e$-folds left to the end of inflation, namely
\begin{equation}
N=\log\left[\frac{a(t_0)}{a(t)}\right]\,,\label{N}
\end{equation}
with $a(t_0)$ the scale factor at the time $t_0$ when acceleration finishes. By recasting this expression in  Equations~(\ref{EOM1}, \ref{conslaw}) and by taking into account that $dN=-Hdt$,
we get
\begin{equation}
3H^2 +9\tilde\alpha H^4\phi'^2=\rho(\phi, X)+\tilde\beta H^2\frac{\phi'^2}{2}\,,\label{EOM1bis}
\end{equation}
\begin{equation}
-\rho'(\phi, X)+3H^2\phi'^2(p_X(\phi, X))
=H^2\phi'\phi''(\tilde\beta-6\tilde\alpha H^2)+H H'\phi'^2(\tilde\beta-18\tilde\alpha H^2)
-3 H^2\phi'^2 (\tilde\beta-6\tilde\alpha H^2)\,.\label{conslawbis}
\end{equation}
Here, the prime denotes the derivative with respect to $N$ and for simplicity we posed
\begin{equation}
\tilde\alpha=\gamma-\alpha\,,\quad\tilde\beta=-2\beta\,.
\end{equation}
Moreover, one has $X=H^2\phi'^2/2$. Given a de Sitter expansion, the field must move very slowly and we may assume $H^2\phi'^2\ll |1/\tilde\alpha|$, such that 
Eqs.(\ref{EOM1bis}, \ref{conslawbis}) read, in the slow-roll regime with $|H'/H|\ll 1$ and $|\phi''|\ll  |\phi'|$,
\begin{equation}
3H^2\simeq \rho(\phi, X)\,,\quad
\rho'(\phi, X)-3H^2 \phi'^2 p_X(\phi, X)\simeq 3H^2\phi'^2(\tilde\beta-6\tilde\alpha H^2)\,.\label{eqslowroll}
\end{equation}
At the beginning of accelerated phase, the field is negative and its magnitude and energy density very large, while at the end, when $N\rightarrow 0$, they tend to vanish. Thus, we have
$\phi'<0$ and $0<\rho'(\phi, X)$, while $X$ increases when $N$ decreases to permit a graceful exit from inflation.

During inflation, 
the slow-roll parameter
\begin{equation}
\epsilon=\frac{H'}{H}\,,\label{epsilon}
\end{equation}
is positive and small. Inflation ends when $\epsilon=1$ and must lead to a total $e$-folds 
$\mathcal{N}\equiv N(a(t_\text{i}))$, where $t_\text{i}$ is the initial time of acceleration, large enough to explain the thermalization of observable universe, namely $55<\mathcal N<65$. 

Let us see some examples of $k$-essence models for inflation.

\subsection{$k$-essence with $p(\phi, X)=F(X)-V(\phi)$}

A suitable form of $k$-essence for inflation is given in the form,
\begin{equation}
p(\phi, X)=F(X)-V(\phi)\,,
\end{equation}
where $F(X)$ and $V(\phi)$ are two functions depending on $X$ and $\phi$, separately. The mechanism for the early-time acceleration is the following: the potential $V(\phi)$ supports the de Sitter expansion as long as the magnitude of $\phi$ is almost a constant, while the 
``kinetic'' part $F(X)$ makes inflation to end when $\phi'^2$ increases.

For example, we may consider,
\begin{equation}
F(X)=\mu X^\lambda\,,\quad 0<\mu\,, \quad 1\leq\lambda\,,
\end{equation}
where $\lambda\,,\mu$ are positive parameters and in the limits $\mu=\lambda=1$ we recover the case of canonical scalar field. From (\ref{rhophi}) we get
\begin{equation}
p(\phi, X)=\mu X^\lambda-V(\phi)\,,\quad \rho(\phi, X)=\mu(2\lambda-1) X^\lambda+V(\phi)\,.
\label{21}
\end{equation}
If we require that 
\begin{equation}
X^\lambda \ll V(\phi\rightarrow-\infty)\,,\quad V(\phi\rightarrow 0^-)\ll X^\lambda\,,
\end{equation} 
the solutions of (\ref{eqslowroll}) are derived as
\begin{equation}
H^2=\frac{V(\phi)}{3}\,,\quad \phi'\simeq\frac{V_\phi(\phi)}{3H^2\left[\mu\lambda(H^2\phi'^2/2)^{\lambda-1}+\tilde\beta-6\tilde\alpha H^2\right]}\,.
\end{equation}
If $\lambda=\mu=1$, one finds the case analyzed in Ref.~\cite{mioH} for canonical scalar field in Horndeski gravity. If $1<\lambda$, we can avoid the contribute of $(H^2\phi'^2)^{\lambda-1}\ll 1$ in the second expression above. In this case, if $\tilde\beta\neq 0$, the analysis of the model is the same of the one of canonical scalar field in Horndeski theory after the redefinition $\tilde\beta\rightarrow 1+\tilde\beta$. On the other hand, if $\tilde\beta=0$, we have
\begin{equation}
\epsilon=-3\tilde\alpha\phi'^2 H^2\,,\quad \tilde\alpha<0\,.
\end{equation}
For example, to obtain
\begin{equation}
H^2=H_0^2(N+1)\,,\quad \epsilon=\frac{1}{2(N+1)}\,,\label{25}
\end{equation}
with $H_0$ the Hubble parameter at the end of inflation, we need
\begin{equation}
\phi'=-\sqrt{-\frac{1}{6\tilde\alpha}}\frac{1}{H_0(N+1)}\,,\quad \phi=\phi_0-\sqrt{-\frac{1}{6\tilde\alpha}}\frac{1}{H_0}\log[N+1]\,,
\end{equation}
$\phi_0$ being the value of the field at the end of inflation. As a consequence, the potential is given by
\begin{equation}
V(\phi)=3H_0^2\text{e}^{H_0(\phi_0-\phi)\sqrt{-6\tilde\alpha}}\,,
\end{equation}
where we remember $\phi<0$.

\subsection{$k$-essence with $p(\phi, X)= g(\phi) X^\lambda-V(\phi)$}

Let us assume the following form of Lagrangian for $k$-essence,
\begin{equation}
p(\phi, X)=g(\phi)X^\lambda-V(\phi)\,,\quad 1\leq\lambda\,,
\end{equation}
where $g(\phi)$ and $V(\phi)$ are functions of the field an $\lambda$ is a positive number. For $g(\phi)=\mu$, we recover the case of the preceding subsection. Now the energy density of $k$-essence is given by
\begin{equation}
\rho(\phi, X)=(2\lambda-1)g(\phi)X^\lambda+V(\phi)\,.\label{29}
\end{equation}
Inflation is realized under the conditions 
\begin{equation}
g(\phi\rightarrow\infty)X^\lambda\ll V(\phi\rightarrow\infty)\,,\quad
V(\phi\rightarrow 0^-)\ll g(\phi\rightarrow 0^-)X^\lambda\,,
\end{equation}
and
the solutions of (\ref{eqslowroll}) are 
\begin{equation}
H^2=\frac{V(\phi)}{3}\,,\quad \phi'\simeq\frac{V_\phi(\phi)}{3H^2\left[\lambda g(\phi)X^{\lambda-1}+\tilde\beta-6\tilde\alpha H^2\right]}\,.\label{27}
\end{equation}
Thus, some interesting configurations can be found. For the special choice $\lambda=1$, one obtains a fluid model with standard kinetic term coupled with the field. 
In this case, for
\begin{equation}
g(\phi)=2\tilde\alpha V(\phi)\,,\quad \lambda=1\,,
\end{equation}
and $\tilde\beta=1$, one recover the results of chaotic inflation,
\begin{equation}
\phi'\simeq\frac{V_\phi(\phi)}{3H^2}\,,\quad \tilde\beta=1\,.
\end{equation}
Otherwise, if $g(\phi)=\mu V(\phi)$, $\mu\neq 2\tilde\alpha$ being a generic parameter, the behavior of the model turns out to be the one of simple canonical scalar field in Horndeski gravity analyzed in Ref.~\cite{mioH},
\begin{equation}
\phi'\simeq\frac{V_\phi(\phi)}{3H^2\left(\tilde\beta+3H^2(\mu-2\tilde\alpha)\right)}\,.
\end{equation} 
When $1<\lambda$, if we avoid the contributes of higher corrections of Horndeski Lagrangian, namely,
\begin{equation}
|\tilde\beta-6\tilde\alpha H^2|\ll\lambda g(\phi) X^{\lambda-1}\ll\frac{3H^2}{X}\,,
\label{31}
\end{equation}
we obtain from (\ref{27}),
\begin{equation}
\phi'\simeq\left[\frac{2^{\lambda-1}V_\phi(\phi)}{3H^{2\lambda}\lambda g(\phi)}
\right]^{\frac{1}{2\lambda-1}}\,.
\end{equation}
In this context, we may consider
\begin{equation}
V(\phi)=\nu\frac{(-\phi)^n}{n}\,,\quad g(\phi)=\mu(-\phi)^m\,,\quad 0< n,m\,,\label{exex}
\end{equation}
where $0<\mu\,,\nu$ are dimensional positive parameters and $m, n$ positive numbers. Therefore, we get
\begin{equation}
\phi'\simeq -\xi(-\phi)^\zeta\,,\quad
\phi\simeq -((1-\zeta)\xi N)^{1/(1-\zeta)}\,,
\end{equation}
where
\begin{equation}
\xi=\left[\frac{6^{\lambda-1}n^\lambda}{\nu^{\lambda-1}\lambda\mu}\right]^{\frac{1}{2\lambda-1}}\,,\quad
\zeta={\frac{n-1-n\lambda-m}{2\lambda-1}}\,,
\end{equation}
and we can verify that, given $1\leq \lambda$, condition (\ref{31}) may be satisfied under the additional restriction $2-m/(\lambda-1)<n$. Note that $\zeta<0$, such that $\phi'\ll 1$ and $\phi\ll 0$ during inflation. Finally, the Hubble parameter results to be
\begin{equation}
H^2\simeq\frac{\nu}{3n}\left((1-\zeta)\xi N\right)^{\frac{n}{1-\zeta}}\,,\quad
\epsilon\simeq\frac{n}{2N(1-\zeta)}\,.
\label{40}
\end{equation}
In considering higher corrections of Horndeski Lagrangian we will pose $\lambda=3/2$, such that from (\ref{27}) we can easily derive
\begin{equation}
\phi'\simeq
-\frac{\sqrt{2} \left(\sqrt{H-\sqrt{2} g V_\phi}-\sqrt{H}\right) \left(\tilde\beta -6
   \tilde\alpha  H^2\right)}{3 g H^{3/2}}
\simeq\frac{V_\phi}{3H^2}(\tilde\beta-6\tilde\alpha H^2)\,,
\end{equation}
where $0<\tilde\beta-6\tilde\alpha H^2$ and we assume $|g(\phi)V_\phi|\ll H$. As an example, we may take 
\begin{equation}
V(\phi)=\nu\frac{(-\phi)^2}{2}\,,
\end{equation}
where $0<\nu$ is positive constant and $g(\phi)\propto1/(-\phi)$. In this case, the kinetic term of $k$-essence results to be proportional to $(-\phi)^2(-\phi')^3$, namely $g(\phi)X^{3/2}\propto (-\phi)^2(-\phi')^3$. We have
\begin{equation}
\phi'\simeq\frac{2(\tilde\beta-\nu\tilde\alpha\phi^2)}{\phi}
\quad\phi\simeq-\sqrt{\frac{\tilde\beta}{\tilde\alpha\nu}}\sqrt{1-\text{e}^{-\tilde\alpha\nu N}}\,,
\end{equation}
such that
\begin{equation}
H^2\simeq\frac{\tilde\beta\left(1-\text{e}^{-\nu\tilde\alpha N}\right)}{6\tilde\alpha}\,,\quad
\epsilon\simeq
\frac{1}{2\left(\text{e}^{\tilde\alpha\nu N}-1\right)}\,,
\end{equation}
and we see that $\epsilon\ll 1$ for large values of $N$.

\subsection{Reconstruction of $k$-essence models for inflation}

The Lagrangian of $k$-essence leads to a huge variety of models to reproduce inflation. In this subsection, we will see how it is possible to reconstruct the form of the Lagrangian by starting from a given solution. 
We note  
that the general reconstruction technique in this class of models has been carried out in Ref.~\cite{BambaGal}, where the method has been applied for general solutions\footnote{In Ref.~\cite{BambaGal} the authors express the scale factor as $a(t)=\exp[g(t)]$, where $g(t)$ is a function of the cosmological time such that $H(t)=\dot g(t)$. In our case, we express the results in terms of the $e$-folds $N$ in (\ref{N}). For a comparison, one must pose $N=g(t_0)-g(t)$.}.

Let us consider the following behavior of the Hubble parameter,
\begin{equation}
H^2=H_0^2(N+1)\,,\quad \epsilon=\frac{1}{2(N+1)}\,,\label{lor0}
\end{equation}
where the constant $H_0$ represents its value at the end of inflation. Thus, from (\ref{eqslowroll}) we have
\begin{equation}
\rho(N)\simeq 3H_0^2(N+1)\,,\quad
p_X(N)=\frac{1+(1+N)(6H_0^2\tilde\alpha(1+N)-\tilde\beta)\phi'^2}{(1+N)\phi'^2}\,,
\label{rsol1}
\end{equation}
with $\rho(\phi, X)\equiv\rho(N)\,,p(\rho,\phi)\equiv p(N)$. In this case,we may assume
\begin{equation}
\phi=\phi_0(1+N)\,,\quad X=\frac{H_0^2}{2}(1+N)\phi_0^2\,,\label{40}
\end{equation}
where $\phi_0<0$ is the value of the field at the end of the early-time acceleration. Thus, the pressure of $k$-essence could have the form
\begin{equation}
p(\phi, X)=-\tilde\beta X+\frac{6\tilde\alpha X^2}{\phi_0^2}+\frac{H_0^2}{2}\log[X/X_0]-V(\phi)\,,
\end{equation}
$X_0\ll 1$ being the kinetic energy of $k$-essence at the end of inflation, 
namely $X_0=H_0^2\phi_0^2/2$,
and $V(\phi)$ a function of the field only fixed by (\ref{rhophi}) and the first equation in (\ref{rsol1}).
In our case,
\begin{equation}
V(\phi)=\frac{H_0^2}{2}
\left(
4-9H_0^2\tilde\alpha\phi_0^2-9H_0^2 N^2\tilde\alpha\phi_0^2+
\tilde\beta\phi_0^2+
N(6-18H_0^2\tilde\alpha\phi_0^2+\tilde\beta\phi_0^2)+
\log[1+N]
\right)\,,
\end{equation}
and by using (\ref{40}) one finally derives
\begin{eqnarray}
p(\phi, X)&=&-\tilde\beta X+\frac{6\tilde\alpha X^2}{\phi_0^2}+\frac{H_0^2}{2}\log[X/X_0]\nonumber\\
&&
\hspace{-4cm}-
\frac{H_0^2}{2}
\left(
4-9H_0^2\tilde\alpha\phi_0^2-9H_0^2 (\phi/\phi_0-1)^2\tilde\alpha\phi_0^2+
\tilde\beta\phi_0^2+
(\phi/\phi_0-1)(6-18H_0^2\tilde\alpha\phi_0^2+\tilde\beta\phi_0^2)+
\log[\phi/\phi_0]
\right)\,.\label{lor1}
\end{eqnarray}
Other possibilities are allowed. 
For example, from (\ref{40}) we can also require
\begin{equation}
p(\phi, X)=\frac{X(1+\phi\phi_0(6\tilde\alpha H_0^2(\phi/\phi_0)-\tilde\beta))}{\phi\phi_0}
-V(\phi)\,,\label{lor2}
\end{equation}
such that
\begin{equation}
V(\phi)=-\frac{H_0^2(\phi/\phi_0)(\phi_0+6\tilde\alpha H_0^2\phi^2\phi_0-\phi(6+\tilde\beta\phi_0^2))}{2\phi}\,.
\end{equation}
One may also pose
\begin{equation}
p_X(\phi, X)=1-V(\phi)\,,
\end{equation}
with
\begin{equation}
\phi'=-\left(
1+N-6H_0^2\tilde\alpha-12\tilde\alpha H_0^2 N
-6\tilde\alpha H_0^2 N^2+\tilde\beta+\tilde\beta N
\right)^{-1/2}\,,
\end{equation}
and one finds the result of Ref.~\cite{mioH} for canonical scalar field models.

A generalization of the preceding example is given by (\ref{lor0}) with
\begin{equation}
\phi=\phi_0(1+N)^n\,,\quad 0<n\,,\label{gen}
\end{equation}
$\phi_0<0$ being a negative constant. This solution can be realized (in the limit $1\ll N$) for the following $k$-essence model,
\begin{equation}
p(\phi, X)=
X\left(6\tilde\alpha H_0^2(\phi/\phi_0)^{1/n}-\tilde\beta+\frac{(\phi/\phi_0)^{1/n-2}}{n^2\phi_0^2}\right)\,,\label{566}
\end{equation}
with
\begin{equation}
V(\phi)\simeq-
\frac{1}{2}H_0^2 n^2 \phi^2\left(\frac{\phi}{\phi_0}\right)^{-1/n}
\left(-\tilde\beta+6\tilde\alpha H_0^2\left(\frac{\phi}{\phi_0}\right)^{1/n}\right)
\,.
\end{equation}
Let us now have a look for the (quasi)-de Sitter solution
\begin{equation}
H^2=H_\text{dS}^2\left[
1-\frac{1}{(N+1)}
\right]^2\,,\quad
\epsilon\simeq\frac{1}{(N+1)^2}\,,\label{56}
\end{equation}
where $H_\text{dS}$ is the constant de Sitter parameter during inflation, when $H\simeq H_\text{dS}$ at $1\ll N$.
By assuming
(\ref{gen})
 again, we may derive the following form of the $k$-essence,
\begin{equation}
p(\phi, X)=X\left(6\tilde\alpha H_\text{dS}^2-\tilde\beta+\frac{2}{\phi^2 n^2}\right)-V(\phi)\,,
\label{57}
\end{equation}
with
\begin{equation}
V(\phi)\simeq
-\frac{H_\text{dS}^2\phi_0^2}{2\phi^2}(2+6\tilde\alpha H_\text{dS}^2 n^2\phi^2-\tilde\beta n^2\phi^2)
\left(\frac{\phi}{\phi_0}\right)^{2-2/n}\,.
\end{equation}
It is understood that other reconstructions are permitted.

\section{Cosmological bounce}

As an alternative description to the Big Bang theory, one may consider the cosmological bounce, where, instead from an initial singularity, the expanding universe emerges from a preceding contracting phase (see Ref.~\cite{Novello} and reference therein for review). The Hubble parameter reads
\begin{equation}
H=h_0(t-t_0)^{2n+1}\,,\quad 0<h_0\,,\label{Hbounce}
\end{equation} 
with $h_0$ a positive dimensional constant, $n$ a natural number, and $t_0$ the fixed time of the bounce. Thus, for $t<t_0\,, H<0$ we have a contraction, while for $t_0<t\,, 0<H$ we have an expansion.

We will study the bounce in terms of the $e$-folds (\ref{N}), where now $a (t_{0})$ is the scale factor at the time of the bounce. With this prescription, $N$ is negative defined, being $a(t_0)$ the minimal value of $a(t)$.
Since
\begin{equation}
a(t)=a(t_0)\exp\left[\frac{h_0(t-t_0)^{2+2n}}{(2+2n)}\right]\,,
\end{equation}
we obtain
\begin{equation}
N=
-\frac{h_0(t-t_0)^{2+2n}}{2+2n}\,.
\end{equation}
From (\ref{Hbounce}) we have
\begin{equation}
|H|=\tilde h_0(-N)^q\,,\quad \tilde h_0=h_0^{\frac{1}{2+2n}}(2+2n)^{\frac{1+2n}{2+2n}}\,,
\quad q=\frac{1+2n}{2+2n}\,.
\end{equation}
Some simple solutions 
can be inferred from (\ref{EOM1bis}\,, \ref{conslawbis}) in order to reconstruct the $k$-essence models realizing the bounce. Given a specific bounce solution, one can find $\rho(\phi, X)\equiv\rho(N)$ from (\ref{EOM1bis}) and therefore plug it inside (\ref{conslawbis}). Thus, the Lagrangian of $k$-essence can be reconstructed 
for simple behaviors of the field.

As an example, we take the simplest case $q=1/2$, namely $n=0$ in (\ref{Hbounce}), and we pose $\phi=\phi_0\sqrt{-N}$, $\phi_0$ being a constant, when bounce is realized, such that in general $0<\phi$ and $\phi=0$ at $t=t_0\,,N=0$. In this case Eq.~(\ref{conslawbis}) holds true if
\begin{equation}
p_X(\phi, X)\equiv p_X(N)=-\tilde h_0(\tilde\alpha+6N\tilde\alpha)-\tilde\beta-\frac{4}{\phi_0^2}\,.
\end{equation}
Since $-N=(\phi/\phi_0)^2$, if we want to satisfy also Eq.~(\ref{EOM1bis}), it must be
\begin{equation}
p=\left(-\tilde h_0(\tilde\alpha-6(\phi/\phi_0)^2\tilde\alpha)-\tilde\beta-\frac{4}{\phi_0^2}\right)X
-\frac{\tilde h_0(12\phi^2(2+\tilde h_0^2\tilde\alpha\phi_0^2)+\phi_0^2(4+\tilde h_0^2\tilde\alpha\phi_0^2))}{8\phi_0^2}\,.
\end{equation}
This is the Lagrangian of a model with kinetic term coupled with the field realizing the bounce solution (\ref{Hbounce}) for $n=0$. 

As an other example we consider $q=3/4$, namely $n=1$ in (\ref{Hbounce}). If we assume again $\phi=\phi_0\sqrt{-N}$, $\phi_0$ constant, from Eq.~(\ref{conslawbis}) we derive
\begin{equation}
p_X(N)=-\frac{\tilde h_0^2\sqrt{-N}(5+12N)\tilde\alpha\phi_0^2+2(6+\tilde\beta\phi_0^2)}{2\phi_0^2}\,.
\end{equation}
Thus, by using the fact that $X=\tilde h^2\sqrt{-N}\phi_0^2/8$, we see that the $k$-essence fluid with
\begin{equation}
p=-\frac{X}{\phi_0^2}\left(6+2(5-12(\phi/\phi_0)^2)\tilde\alpha X+\tilde\beta\phi_0^2)
-\frac{3\tilde h_0^2}{32\phi_0^2}\left(
8\sqrt{\frac{\phi^2}{\phi_0^2}}(4\phi^2+\phi_0^2)+\tilde h_0^2\tilde\alpha\phi^2(12\phi^2+5\phi_0^2)
\right)
\right)\,,
\end{equation}
realizes the bounce (\ref{Hbounce}) with $n=1$.

\section{Cosmological perturbations\label{pert}}

In this section, we will discuss perturbations during the early-time acceleration for our class of Horndeski models with $k$-essence.
Scalar perturbations in flat FRW metric (\ref{metric}) reads~\cite{Def, DeFelice},
\begin{equation}
ds^2=-[(1+\alpha(t, {\bf x}))^2-a(t)^{-2}\text{e}^{-2\zeta(t, {\bf x})}(\partial \psi(t,{\bf x}))^2]dt^2+2\partial_i\psi
(t,{\bf x})dt dx^i+a(t)^2
\text{e}^{2\zeta(t, {\bf x})}d{\bf x}\,,
\end{equation}
where $\alpha(t, {\bf x})\,,\psi(t, {\bf x})$ and $\zeta\equiv\zeta(t,{\bf x})$ are functions of space-time coordinates. A direct computation inside the action  (\ref{action}) leads to
\begin{equation}
I=\int_\mathcal{M}dx^4 a^3\left[A\dot\zeta^2-\frac{B}{a^2}(\nabla\zeta)^2\right]\,,\label{pertaction}
\end{equation}
where 
\begin{eqnarray}
&&
A \equiv
\frac{\dot\phi^2(1+\tilde\alpha\dot\phi^2)}{2(H+3H\tilde\alpha\dot\phi^2)^2}
\times\nonumber\\&&
(-6H^2\tilde\alpha+p_X(\phi, X)+\tilde\beta+p_{XX}(\phi, X)\dot\phi^2+\tilde\alpha(18H^2\tilde\alpha+p_X(\phi, X)+\tilde\beta)\dot\phi^2+\tilde\alpha p_{XX}(\phi, X)\dot\phi^4)\,,
\nonumber\\
&&
B \equiv\frac{1}{(H+3H\tilde\alpha\dot\phi^2)^2}\times\nonumber\\
\hspace{-1.5cm}&&\left(
-(1+3\tilde\alpha\dot\phi^2)(-4H^2\tilde\alpha^2\dot\phi^4+\dot H(1+\tilde\alpha\dot\phi^2)^2)
+2H\tilde\alpha\dot\phi(1+\tilde\alpha\dot\phi^2)(-1+3\tilde\alpha\dot\phi^2)\ddot\phi
\right)\,.\nonumber\\&&
\end{eqnarray}
Thus, one has
\begin{equation}
I=\int_\mathcal{M}dx^4 a^3 A\left[\dot\zeta^2-\frac{c_s^2}{a^2}(\nabla\zeta)^2\right]\,,
\label{51}
\end{equation}
where the square of the sound speed is given by
\begin{equation}
\hspace{-2cm}
c_s^2=\frac{
\left(-2(1+3\tilde\alpha\dot\phi^2)(-4H^2\tilde\alpha^2\dot\phi^4+\dot H(1+\tilde\alpha\dot\phi^2)^2)
+4H\tilde\alpha\dot\phi(1+\tilde\alpha\dot\phi^2)(-1+3\tilde\alpha\dot\phi^2)\ddot\phi\right)}
{\dot\phi^2(1+\tilde\alpha\dot\phi^2)\left(-6H^2\tilde\alpha+p_X(\phi, X)+\tilde\beta+
p_{XX}(\phi, X)\dot\phi^2+
\tilde\alpha(18H^2\tilde\alpha+p_X(\phi, X)+\tilde\beta)\dot\phi^2+\tilde\alpha p_{XX}(\phi, X)\dot\phi^4\right)}\,.
\end{equation}
It is necessary to require $0<A, B$ to avoid ghost and instabilities. By introducing the $e$-folds parameter one obtains in the limit $|\dot\phi^2|\ll 1/|\tilde\alpha|$, 
\begin{equation}
c_s^2\simeq\frac{2H'}{(H\tilde\beta+H p_X(\phi, X)-6H^3\tilde\alpha+p_{XX}(\phi, X)H^3\phi'^2)\phi'^2}\,.
\end{equation}
By using the conservation law in (\ref{eqslowroll}) one immediatly has that in the case of canonical scalar field with $p_{XX}(\phi, X)=0$, $c_s^2=1$. However, for $k$-essence and higher order kinetic term with $0<p_{XX}$, the sound speed results to be smaller than one since $c_s^2\simeq2H'/(2H'+p_{XX}H^3\phi'^4)$.

By substituting 
\begin{equation}
v\equiv v(t, {\bf x})=z(t) \zeta(t, {\bf x})\,,\quad z\equiv z(t)=\sqrt{a^3 A}\,,
\end{equation}
from (\ref{51}), after integration by part, we obtain
\begin{equation}
I=\int dx^4\left[\dot v^2-\frac{c_s^2}{a^2}(\nabla v)^2+\ddot z\frac{v^2}{z}\right]\,,\label{action2}
\end{equation}
which leads to
\begin{equation}
\ddot v-\frac{c_s^2}{a^2}\bigtriangleup v-\frac{\ddot z}{z}v=0\,.
\end{equation}
By decomposing $v(t, {\bf x})$ in Fourier modes $v_k\equiv v_k(t)$ whose explicit dependence on $\bf k$ is given by $\exp[i {\bf k}{\bf x}]$, we get
\begin{equation}
\ddot v_k+\left(k^2\frac{c_s^2}{a^2}-\frac{\ddot z}{z}\right)v_k=0\,.
\end{equation}
The short-wave solution of this equation for $1\ll k^2/a^2$ is given by
\begin{equation}
v_k\simeq c_k \text{e}^{\pm i k\int \frac{c_s}{a}dt}\,, \label{58}
\end{equation}
where $c_k$ is a constant. On the other hand, for long-wave perturbations with $k^2/a^2\ll 1$ we obtain the implicit solution
\begin{equation}
v_k\simeq c_1 z+c_2 z\int\frac{dt}{z^2}\,,\label{59}
\end{equation}
with $c_1\,,c_2$ integration constants. The explicit solution for perturbations in quasi de-Sitter space-time can be derived as~\cite{DeFelice},
\begin{equation}
v_k(t)\simeq c_0\sqrt{\frac{a}{2}}\frac{a H}{(c_s k)^{3/2}}
\text{e}^{\pm i k\int \frac{c_s}{a}dt}
\left(1+i c_s k\int\frac{dt}{a}\right)\,,
\end{equation}
where in the given limits one can recovers (\ref{58})--(\ref{59}) by taking into account that 
$z\simeq\tau^2\sqrt{a/2}$ and $Ha=-1/\tau$, with $\tau=\int dt/a$ the conformal time, and $c_1=0$.
The constant $c_0$ is fixed by the  Bunch-Davies vacuum state 
$v_k(t)=\sqrt{a}\exp[\pm i k\int c_s dt/a]/(2\sqrt{c_s\kappa})$ in the asymptotic past such that $c_0=i/\sqrt{2}$ and finally
\begin{equation}
\zeta_k\equiv \frac{v_k}{\sqrt{A a^3}}=i\frac{H}{2\sqrt{A}(c_s k)^{3/2}}
\text{e}^{\pm i k\int \frac{c_s}{a}dt}
\left(1+i c_s k\int\frac{dt}{a}\right)
\,.
\end{equation}
Thus, the variance of the power spectrum of perturbations on the sound horizon crossing $c_s\kappa\simeq H a$ reads
\begin{equation}
\mathcal P_{\mathcal R}\equiv\frac{|\zeta_k|^2 k^3}{2\pi^2}|_{c_s k\simeq H a}=\frac{H^2}{8\pi^2 c_s^3 A}|_{c_s k\simeq H a}\,.
\end{equation}
As a consequence, the spectral index results to be
\begin{equation}
1-n_s=-\frac{d\ln \mathcal P_{\mathcal R}}{d \ln k}|_{k=a H/c_s}=2\epsilon+\eta_{sF}+s\,,
\end{equation}
with~\cite{DeFelice},
\begin{equation}
\epsilon=-\frac{\dot H}{H^2}\,,\quad \eta_{sF}=\frac{\dot \epsilon_s F+\epsilon_s\dot F}{H (\epsilon_s F)}\,,\quad s=\frac{\dot c_s}{H c_s}\,,\quad \epsilon_s=\frac{A c_s^2}{F}\,,
\end{equation}
and
\begin{equation}
F=1+\alpha\dot\phi^2\,.
\end{equation}
In a similar way, the tensor-to-scalar ratio is derived from the tensor perturbations in flat FRW space-time as
\begin{equation}
r=16 c_s\epsilon_s\,.
\end{equation}
Therefore, in the case of $k$-essence models where $c_s<1$, this quantity may be easily suppressed.
By plugging the $e$-folds (\ref{N}), we obtain for our class of Horndeski models
\begin{eqnarray}
\hspace{-1cm}(n_S-1)&\simeq&\left(
\phi '\left(3 H H'' \left(\tilde\beta +H^2
   \left(p_{XX} \phi '(t)^2-6 \tilde\alpha
   \right)+p_X\right)-H' \left(H' \left(9 H^2
   \left(p_{XX} \phi '^2-6 \tilde\alpha \right)+7 (\tilde\beta
   +p_X)\right)
\right.\right.\right.
\nonumber\\&&
\left.\left.\left.
+H \left(H^2 p_{XX}' \phi'^2+p_X'\right)\right)\right)-2 H H' \phi ''
   \left(\tilde\beta +H^2 \left(2 p_{XX} \phi '^2-6 \tilde\alpha
   \right)+p_X\right)\right)
\nonumber\\&&\times
\frac{1}{2 H H' \phi ' \left(\tilde\beta
   +H^2 \left(p_{XX}\phi '^2-6 \tilde\alpha
   \right)+p_X\right)}\,,
\end{eqnarray}
and
\begin{equation}
r\simeq 16 \sqrt{2} \frac{H'}{H} \sqrt{\frac{H'}{H \phi '^2
   \left(-\frac{\tilde\beta }{2}+H^2 \left(p_{XX} \phi
   '^2-6 \tilde\alpha \right)+p_X\right)}}\,.
\end{equation}
The last Planck satellite data~\cite{Planck} lead to
$n_{\mathrm{s}} = 0.968 \pm 0.006\, (68\%\,\mathrm{CL})$ and 
$r < 0.11\, (95\%\,\mathrm{CL})$. If we pose $N\equiv \mathcal N\simeq 60$, the scenario is viable if $(n_s-1)\simeq -2/N$.

When $p_X=1$ (canonical scalar field) and $\tilde\alpha=\tilde\beta=0$ (chaotic inflation) we obtain
\begin{equation}
(n_s-1)\simeq-\frac{7H'}{2H}+\frac{3H''}{2H'}-\frac{\phi''}{\phi'}\,,\quad
r\simeq16\sqrt{2}\phi'^2\left(\frac{H'}{H\phi'^2}\right)^{3/2}\,,
\end{equation}
or, in terms of the cosmological time, by using equations (\ref{eqslowroll}),
\begin{equation}
(n_s-1)\simeq\frac{4\dot H}{H^2}-\frac{\ddot H}{\dot H H^2}\,,\quad r\simeq -16\frac{\dot H}{H^2}\,,
\end{equation}
which correspond to the usual relations for chaotic canonical scalar field inflation.
In the case of model (\ref{21}) for solution (\ref{25}), the spectral index reads $(n_s-1)\simeq-7/(2(N+1))$, while for the tensor-to-scalar ratio one can see that $r\propto (N+1)^{3/2}$. 
In the case of model (\ref{lor1}) for solution (\ref{lor0}), the spectral index results to be $(n_s-1)\simeq-9/(2(N+1))$, while for the tensor-to-scalar ratio one finds
$r\propto (1+N)^{-5/2}$. It means that this models predict a total e-folds extremely large ($100<N$) 
in order to be in agreement with the cosmological observations.

For the class of models (\ref{29}), we may consider the example in (\ref{exex}) with $\lambda=2$ and $m=1$. Given the solution (\ref{40}), we find
\begin{equation}
(n_s-1)\simeq-\frac{(11+10n)}{2(5+n)(N+1)}
\end{equation}
Thus, if we assume $N\simeq 60$, the model is in agreement with Planck data when $n=3/2$. In this case the tensor-to-scalar ratio,
\begin{equation}
r\simeq\frac{54\left(\frac{3}{13}\right)^{4/13}2^{21/26}}{13 N^{17/13}}\sqrt{-\frac{1}{\tilde\alpha\nu\xi^{21/13}}}\,,
\end{equation}
is small enough and the model is viable. Note that the given values of $\lambda, m$ and $n$ bring to satisfy also condition (\ref{31}).

The class of models (\ref{566}) or (\ref{lor2}) for the specific case $n=1$, which admit the inflationary solution (\ref{lor0}), leads to
\begin{equation}
(n_s-1)\simeq -\frac{2}{N}\,\quad r\simeq\frac{8}{N}\,,
\end{equation}
like in the case of canonical scalar field with quadratic potential in the background of General Relativity. Here, the tensor-to-scalar ratio is slightly larger respect to the observed one. 

Finally, for the model in (\ref{56})--(\ref{57}), we find
\begin{equation}
(n_s-1)\simeq -\frac{2+n}{N}\,,\quad r\simeq
8\sqrt{\frac{2}{3}}\sqrt{\frac{1}{H_\text{dS}^2 n^2\tilde\alpha\phi_0^2}}\frac{1}{N^{n+3/2}}\,.
\end{equation}
Thus, by considering $N\simeq 60$, this theory is viable for small values of $n$.

\section{Conclusions}

In this paper, we investigated cosmological solutions for early-time universe in a simple class of Horndeski models where the scalar field plays the role of $k$-essence. The Lagrangian of the theory is quite involved and higher curvature corrections to General Relativity emerge at large curvatures contributing to the primordial acceleration. The advantage
of Horndeski models is that the field equations are at the second order like in General Relativity. 
Moreover, $k$-essence is one of the possible descriptions for inflation and its
Lagrangian includes higher order kinetic term and may describe a huge variety of effective fluids, appearing as a quite general theory (see also Ref.~\cite{ultimok}). 

Solutions for early-time acceleration have been investigated. In this respect, we considered different forms of Lagrangian for $k$-essence and we used a reconstruction method to infer Lagrangian by starting from the given solutions. Since a model for inflation must lead to a correct amount of $e$-folds and a correct prediction for the spectral index and the tensor-to-scalar ratio, we furnished the formalism to calculate perturbations and we applied the results to our examples. A section has been devoted to the study of cosmological bounce as an alternative description to the Big Bang. $k$-essence in Horndeski gravity admitting bounce solutions has been reconstructed.

For bouncing cosmology in modified gravity see also Ref.~\cite{Odbs, Odbs2, Odbs3}. For some reviews of inflation in modified gravity see Ref.~\cite{Odinfrev, myrevinfl}. Since a special class of Horndeski models can be recasted in the form of Gauss-Bonnet modified gravity, it may be useful to compare our results with the ones in Ref.~\cite{RGinfl}. Other useful references can be found in Refs.~\cite{uno, unobis, due, tre}.


\end{document}